\documentclass[aps,prb,superscriptaddress,twocolumn,floatfix,citeautoscript]{revtex4-2}
\usepackage{graphicx}
\usepackage{dcolumn}
\usepackage{bm}
\usepackage{xcolor}
\usepackage{float}
\usepackage[colorlinks = true,
            linkcolor = blue,
            urlcolor  = blue,
            citecolor = blue,
            anchorcolor = blue]{hyperref}
\usepackage{graphicx,latexsym} 
\usepackage{amssymb,amsmath}
\usepackage{amsthm}
\usepackage{longtable,booktabs,setspace}

\usepackage{url}
\DeclareMathOperator{\sech}{sech}
\usepackage{graphicx} 
\usepackage{url}

\newcommand{\NM}[1]{\textcolor{black}{ #1 }}

\begin{document}
\title{An activation-relaxation technique study of two-level system impact on internal dissipation using DFT-based moment tensor potential}
\author{Renaude Girard}
\email{renaud.girard@umontreal.ca}
\affiliation{%
Département de physique, Institut Courtois and Regroupement québécois sur les matériaux de pointe,
Université de Montréal, C.P. 6128, Succursale Centre-Ville,
Montréal, H3C~3J7, Québec, Canada.\\
}%
\author{Carl Lévesque}
\author{Normand Mousseau}
\email{normand.mousseau@umontreal.ca}
\affiliation{%
Département de physique, Institut Courtois and Regroupement québécois sur les matériaux de pointe,
Université de Montréal, C.P. 6128, Succursale Centre-Ville,
Montréal, H3C~3J7, Québec, Canada.\\
}%
\author{François Schiettekatte}
\email{francois.schiettekatte@umontreal.ca}
\affiliation{%
Département de physique, Institut Courtois and Regroupement québécois sur les matériaux de pointe,
Université de Montréal, C.P. 6128, Succursale Centre-Ville,
Montréal, H3C~3J7, Québec, Canada.\\
}%
\begin{abstract}\label{abstract}
We use a  recently-developed machine-learned Moment Tensor Potential (MTP) trained on data generated with the density functional theory (DFT) and tailored to amorphous silicon  coupled with the Activation-Relaxation Technique nouveau (ARTn) to identify and classify two-level systems (TLS). The samples generated using MTP recover experimental results and provide  average structural and dissipative properties similar to those obtained with a modified Stillinger-Weber potential, including radial distribution function, defect concentration and internal friction. Atomistic details, however, are significantly different, including the density and type of TLS. In particular, we find that while the density of TLS involving a bond-hopping mechanism is similar for the two potentials,  more complex TLSs, such as those involving a Wooten-Winer-Weaire  bond exchange, are about twice as common. Analysis also shows that TLSs, for MTP-based models, are mostly isolated and oscillate independently from each other.
\end{abstract}
\maketitle
\section{Introduction}\label{introduction}

After ten years of operation since the first detection, large-scale gravitational wave detectors (GWD) have led to the observation of hundreds of gravitational waves~\cite{gwtc40introductionversion40}, confirming so far Einstein's general theory of gravitation~\cite{clesse_clustering_2017}. These detectors are based on a Michelson interferometer with arms a few kilometers long each containing a Fabry-Perot cavity~\cite{abbott_gw150914_2016}. With current detectors, events involving compact objects, for example, the collision of two black holes, can be identified reliably. However, the detection of events involving smaller or more distant masses, with sufficient precision to characterize them is limited by the background noise. In order to access and characterize those events, considerable efforts were spent in the comprehension and identification of the sources of the background noise. 

For current GWD experiments, the sensitivity is optimal around 100\,Hz \cite{nawrodt_challenges_2011}. At this frequency, the main sources of noise come from fluctuations in mirror coating and quantum noise\cite{waldman_advanced_2011}. Analysis based on the fluctuation-dissipation theory \cite{kubo_fluctuation-dissipation_nodate} shows that the coating thermal noise originates from microscopic mechanical fluctuations around local configurational minima, which can be attributed to classically activated atomic-scale two-level systems (TLS) \cite{phillips_two-level_1987}. TLSs correspond to systems composed of two close-by local metastable states of similar energy separated by a relatively low-energy barrier. Although the general description of TLSs was provided many decades ago, the mechanistic details, which may depend closely on the specificity of the materials, are still lacking, hence a handle on ways to control them. Special focus has been given to TiO$_2$:Ta$_2$O$_5$~\cite{PhysRevB.93.014105,Prasai2023} and SiO$_2$, the former being the high index material in the current GWD reflective stacks and is the one contributing the most to the coating thermal noise, while the latter is the low index material and is one of the amorphous materials with the lowest internal dissipation at room temperature~\cite{Harry_2002} .

Amorphous silicon (\textit{a}-Si) also features a low internal dissipation~\cite{Hellman_PRL2014} hence low noise while it has a high refractive index, and its incorporation in optical stacks for GWD is a subject of active research~\cite{PhysRevD.91.042001,MolinaRuiz2024,Lalande2025}. Recently,  atomistic mechanisms responsible for the mechanical loss associated with TLSs were identified in models of \textit{a}-Si described with the empirical modified Stillinger-Weber potential (mSW) \cite{vink_fitting_2001} using an open-ended transition state search method, the \emph{Activation-Relaxation Technique nouveau} (ARTn)\cite{levesque_internal_2022}. As \textit{a}-Si is a  mono-atomic amorphous material and a potential candidate for the next generation of GWD mirrors~\cite{Birney_2018}, this characterization offered both generic information on the atomistic origin of TLSs and specific details on dissipation mechanisms in a potentially relevant material. While this study provided a coherent picture of TLSs that is in general agreement with experimental dissipation measurements, it suffers from the use of an empirical potential that was fitted to reproduce the overall structural properties of \textit{a}-Si, but not its dynamical properties. Indeed, as shown, for example, in Ref.~\cite{jay_simulation_2018}, transition-state configurations generated with this potential in an ion-implanted crystalline silicon simulation box may not correspond to barriers predicted with density functional theory (DFT). 

Since the concentration of TLSs in \textit{a}-Si is low, their study requires accumulating considerable statistics on simulation cells too large to be explored using density functional theory (DFT) approaches. With recent advances in machine learning, it is now possible to circumvent this barrier using rich machine-learned potential (MLP) that are fitted with high precision to DFT calculations~\cite{novikov_improving_2019}. Contrary to DFT, these MLPs are localized in nature and, therefore, scale well with system size while providing a precise physical description at accessible computational costs. In this work, we reexamine the fundamental nature of the atomistic mechanisms responsible for TLSs in \textit{a}-Si using a recently developed high-quality universal MLP for Si based on the moment-tensor potential formalism\cite{shapeev_moment_2016} fitted to DFT calculations~\cite{zongo_unified_2024}. We also compare our results with both experiments and recent modeling efforts using a purely empirical potential, which allows us to compare a well-established empirical potential directly to an MLP in the context of a complex structure. We identify subtle but notable differences in the geometrical details of TLSs that might impact strategies to handle those for gravitational wave detection, for example. 

This paper is constructed as follows. We first present a brief overview of TLS before describing the modelling approach and the analysis methods. We then present results and compare with TLSs generated with empirical potentials, showing that the MLP-based models find twice as many defects as the empirical potential, although the mechanisms identified are more compact and complex. 

\section{Theory and Methods}\label{theory}

This section provides a short introduction to TLS and to the generation and analysis methods used to explore TLSs in realistic models of \textit{a}-Si. 

\subsection{Two-level systems}\label{tls}
    
The time evolution of a mechanical system can be described as a series of thermally activated events. Each event involves a transition from a metastable state to another through a trajectory characterized by an energy barrier defined at the 
first-order transition state. While, in disordered systems, these energy barriers can be associated with a broad energy spectrum, events contributing to the background noise in GWD correspond to a  small subset of these events, associated with TLSs, which are characterized by oscillations between two states of similar energy with a barrier corresponding to frequencies between 10 and 10~000\,Hz. 

\begin{figure}
    \centering
    \includegraphics[width=1.0\linewidth]{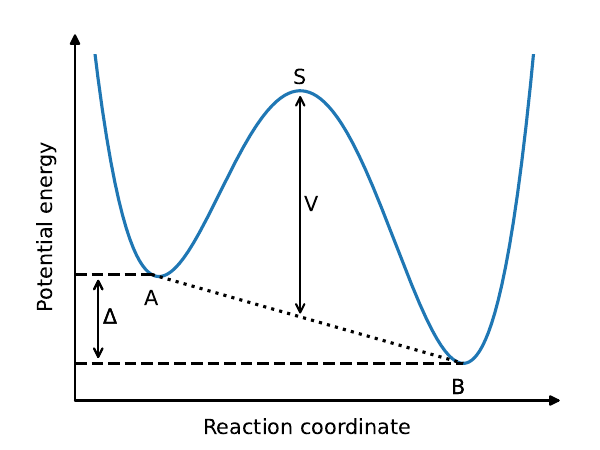}
    \caption{Representation of the configurational energy of a TLS as a function of its reaction coordinate. This TLS consists of two minima, $A$ and $B$, separated by a saddle point $S$ surrounded by high-energy barriers. $\Delta$ indicates the asymmetry between the two minima and $V$ the mean energy barrier.}
    \label{fig:tls}
\end{figure}

More specifically, a TLS consists of a system of two minima separated by a low-energy single saddle point and surrounded by significantly higher energy barriers, effectively creating an isolated two-state basin (see Fig.~\ref{fig:tls}). 

A TLS is characterized by the energy asymmetry, $\Delta=|E_b - E_a|$, and the mean barrier energy,
\begin{equation}
 V=\frac{E_S-E_a}{2}+\frac{E_S-E_b}{2},
\end{equation}
between the two metastable minima of energy $E_a$ and $E_b$.
%

In the transition state theory \cite{truhlar1996}, we use the Arrhenius law to find the mean rate of transition between two minima in the potential plane. Neglecting quantum tunneling,
\begin{equation}
\label{eq:tst}
         \tau_{i\rightarrow j}=\tau_{0}e^{\frac{E_{\mathrm{sad}}-E_i}{k_bT}},
\end{equation}
\NM{where $\tau_{i\rightarrow j}$ is the average time between transitions from minimum $i$ to minimum $j$, $E_{\mathrm{sad}}$ is the energy at the barrier between $i$ and $j$, $k_b$ is the Boltzmann constant, $T$ is the temperature, and $\tau_{0}^{-1}$ is the attempt frequency of the event.} 

The contribution of TLSs to the background noise can be computed as the loss angle\NM{, which treats each TLS in a symmetric fashion between states $a$ and $b$}: 
\begin{equation}\label{loss_angle_eq}
     Q^{-1}(\omega)=\frac{1}{ \mathcal{V}\,C}\sum_{i ,\mathrm{TLS}}\frac{\gamma^{2}_{i}}{k_bT}\frac{\omega \tau_i}{1+\omega^2\tau^{2}_{i}}\sech^2\left(\frac{\Delta_i}{2k_bT}\right),
\end{equation}
which can be measured experimentally, where $\tau_i$,  the relaxation time, is given by: 
\begin{equation}
         \tau_i=\tau_0\sech\left( \frac{\Delta_i}{2k_bT}\right)e^{\frac{V_i}{k_bT}}.
\end{equation}
Here, $C$ is the elastic modulus (longitudinal or transverse), $\gamma_i$ is the deformation potential, $\mathcal{V}$ the volume of the system.

The full derivation for these equations is given in Ref.~\cite{phillips_two-level_1987}. The coupling tensor $\Bar{\gamma}$, which forms the basis of the deformation potential $\gamma_i$, is a quantity derived from the elastic tensor $\Bar{\epsilon}$, as
\begin{equation}
         \Bar{\gamma}=\frac{\delta \Delta}{\delta \Bar{\epsilon}}.
\end{equation}
Formally,  $\gamma_i$ can be defined in the transverse and the longitudinal directions, according to the direction of the strain. In a disordered system, TLSs are isotropic; as such, we average $\gamma_i$  over all strain directions:
\begin{equation}
\begin{aligned}
\gamma_L^2= & \frac{1}{5}\left(\bar\gamma_{X X}^2+\bar\gamma_{Y Y}^2+\bar\gamma_{Z Z}^2\right) \\
& +\frac{2}{15}\left(\bar\gamma_{X X} \bar\gamma_{Y Y}+\bar\gamma_{X X} \bar\gamma_{Z Z}+\bar\gamma_{Y Y} \bar\gamma_{Z Z}\right) \\
& +\frac{4}{15}\left(\bar\gamma_{X Y}^2+\bar\gamma_{X Z}^2+\bar\gamma_{Y Z}^2\right),
\end{aligned}
\end{equation}
for the longitudinal modulus, and 
\begin{equation}
\begin{aligned}
\gamma_T^2= & \frac{1}{15}\left(\gamma_{X X}^2+\gamma_{Y Y}^2+\gamma_{Z Z}^2\right) \\
& -\frac{1}{15}\left(\gamma_{X X} \gamma_{Y Y}+\gamma_{Y Y} \gamma_{Z Z}+\gamma_{Z Z}\gamma_{ X X}\right) \\
& +\frac{3}{15}\left(\gamma_{Y Y}^2+\gamma_{Y Z}^2+\gamma_{Y Z}^2\right),
\end{aligned}
\end{equation}
for the transverse modulus. The detailed derivation for the equations above can be found in Ref.~\cite{damart_atomistic_2018}. Not all TLSs contribute to noise in the frequency range where GWD detection is affected by this noise. In Ref.~\cite{levesque_internal_2022}, the barrier range of the contributing TLSs   was identified to be between 0.2 and 0.7\,eV.

\subsection{Moment Tensor Potential}\label{mtp}

Recent work on TLSs in amorphous materials was mostly pursued using an empirical potential, whether the van Beest-Kramer-van Santen potential (BKS)~\cite{van_beest_force_1990} for silica ~\cite{damart_atomistic_2018}, a combination of BKS and Morse potential in TiO$_2$:Ta$_2$O$_5$~\cite{PhysRevB.93.014105} or a mSW potential~\cite{stillinger_role_1985,vink_fitting_2001} for \textit{a}-Si~\cite{levesque_internal_2022}.

However, it has been shown that barriers generated with these potentials might not correspond to saddle points obtained with a more precise DFT representation ~\cite{jay_simulation_2018}. As it is not possible to explore broadly with DFT the energy landscape of amorphous systems, we select to use a DFT-based machine-learned potential recently developed for \textit{a}-Si by Zongo \textit{et al.}~\cite{zongo_unified_2024}.

This force field was created using the moment tensor potential description (MTP)~\cite{shapeev_moment_2016} and adjusted to the DFT energies and forces for  a wide  range of silicon structures,  including the amorphous phase.  The theory behind constructing MTP is based on the assumption that it is possible to approximate any potential through a complex series of polynomials.

A more in-depth explanation of MTP and the construction of the polynomial in Ref.~\cite{shapeev_moment_2016}. The parameters used here were developed as part of an effort to construct a universal potential for both Si and SiO$_2$ following a training schedule presented in Ref.~\cite{zongo_unified_2024}.
While the general validity of the potential is presented in that work, a more specific characterization  of the potential's quality for  \textit{a}-Si can be found in Ref.~\cite{zongo_amorphous_2025}. This paper shows that the potential favors low-defect structures with overall properties in excellent agreement with experiments and other high-quality modeling results.

\subsection{Atomic Models}\label{model}

To facilitate the comparison with previously published results, we closely follow the melt-and-quench schedule used in Ref.~\cite{levesque_internal_2022} to generate unbiased and independent high quality 1000-atom \textit{a}-Si models. 

Each model is created starting with a 1000-atom crystalline cell and that is heated at 3000\,K for 1\,ns with a randomly generated initial velocity distribution. The configuration is then brought to 1500\,K at a relatively high rate of $1.5\times 10^{13}$\,K/s and then cooled to 300\,K at a much slower rate of $1.0\times 10^{11}$\,K/s. After a 1\,ns equilibration at 300\,K, to eliminate the low energy  barriers, the configuration is minimized at 0\,K at constant pressure of $P=0$\,Pa. As the computation cost of MTP is about 100 times higher than for the mSW potential due to the high complexity of the local descriptors and the larger number of parameters evaluated as compared  with standard empirical potentials,  we limit the number of independent samples to 28 samples  instead  of the 200 produced with mSW~\cite{levesque_internal_2022}. As seen below, since the density for TLSs in the MTP-based models is higher than for mSW models, the quality of statistics remains sufficient to draw meaningful conclusions.  

\subsection{ART nouveau}\label{artn}

To identify the TLSs present in the system, it is necessary to explore extensively the energy landscape surrounding each final configuration. To do so, we use the Activation-Relaxation Technique nouveau (ARTn)~\cite{barkema_event-based_1996,malek_dynamics,mousseau_activation-relaxation_2012,jay_finding_2020}, an efficient open-ended transition state searching method.
Here, each event search is centered on a specific atom. Starting from a local minimum, an initial random deformation is applied on a central atom and its first neighbors (within a radius of 2.9\,Å). The system is deformed slowly along this direction, with slight perpendicular relaxation at each step to avoid collision, until the lowest eigenvalue of the Hessian matrix becomes negative (below a threshold of -1\,eV/Å), indicating that a direction of instability has been found. The system is then pushed away from the minimum along this direction of negative curvature, with forces minimized in the hyperplane perpendicular to this direction until the total force falls below 0.01\,eV/Å, indicating that a saddle point has been reached. The system is then minimized after being slightly displaced from the saddle point in both directions along the eigenvector corresponding to the negative eigenvalue to find the final state and ensure that the saddle point is connected to the initial state. If this is the case, the event is added to the catalog. \NM{Events are classified according based on the automorphic group associated with the graph formed by bonding network of the atoms found in a radius of XX Å around the central atom. As discussed in Ref.~\cite{mousseau_activation-relaxation_2012}, this graph provides a reliable framework for classifying events.} 

To generate the full event catalog, we launched 30 event searches centered on each atom in the cell, for a total of 30\,000 event searches per configuration, leading to a total of 52\,000 topologically different events with a barrier below 5\,eV  for the 28 configurations.   Only events corresponding to the TLS criteria are discussed below. \NM{As the first event classification done using the local topology can overcount the number of different events, the list of TLSs is further refined with geometrical criteria based on relative displacements} to remove the rare cases of overcounting, when the same structure is linked with two different topologies due to small deformations associated with a soft environment.

\section{Results}

The quality of the resulting configurations is assessed by comparing the structural properties of the MTP generated samples to available experimental data and samples generated through melt and quench using the mSW potential. Events generated are described by a barrier height, the energy asymmetry, and the energy of the original and final minima and of the saddle point. We use the criteria retained in \ref{tls} and described above to identify the TLS that contribute to the internal friction (loss angle) in GWD experiments. The properties of these events is also compared to previous results obtained on samples generated with mSW\cite{levesque_internal_2022}.

\subsection{Prefactor analysis}

Recent work has shown that activated prefactors,$\tau_0$ in Eq.~\ref{eq:tst}, can vary widely in disordered materials \cite{gelin_enthalpy-entropy_2020}. It is therefore necessary to assess whether it is possible to use the general approximation of a constant prefactor for the TLSs considered here. 

To do so, we use the harmonic approximation of the transition state theory (HTST):
\begin{equation}
\tau_{\mathrm{HTST}}^{-1}=\frac{\prod_{i=1}^{3 N} v_i}{\prod_{j=1}^{3 N-1} v_j^{\prime}}
\end{equation}
where $\nu_i$ are the real eigenfrequencies of the Hessian matrix.

Because of the computational cost associated with this calculation, we assess the evolution  of the prefactor  for all 1563 events with a barrier below 5\,eV found in a single representative configuration.  
\begin{figure}[h]
    \centering
    \includegraphics[width=1.0\linewidth]{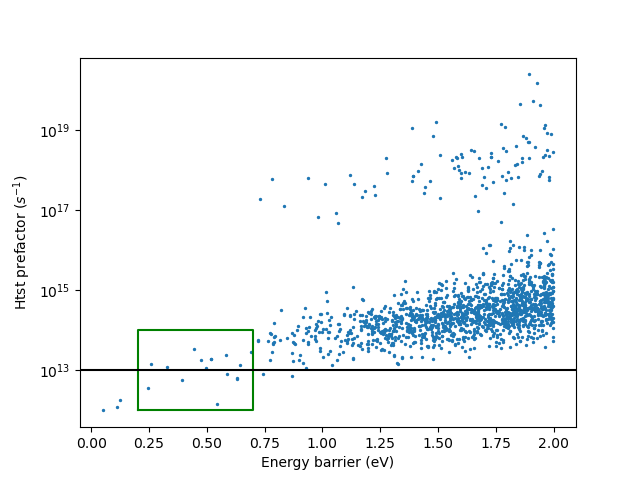}
    \caption{Attempt frequency, or prefactor, given by the harmonic transition theory, as a function of energy barrier for all events with a barrier below 5~eV found in representative 1000-atom \textit{a}-Si configuration prepared as discussed in the text.  The green rectangle indicates the data range studied in the present work and the black line the constant prefactor usually used for \textit{a}-Si }
    \label{fig:prefactor}
\end{figure}
Figure \ref{fig:prefactor} presents the prefactor of each cataloged event as a function of the associated energy barrier of the associated event for this configuration. The green rectangle indicates the region that includes TLSs. While the distribution of rates is broad for events with energy barriers above 1.0\,eV, \NM{and further analysis is required to understand the nature of the very high prefactors observed in this region,}  we see that for energy barriers between 0.2 and 0.7\,eV, the prefactor is found to be  $(1.4\pm 0.9)\times 10^{13}$\,Hz, compatible with the typical constant rate of $1\times 10^{13}$ used previously for computing activation rates~\cite{damart_atomistic_2018,levesque_internal_2022}. It  is therefore appropriate to follow the convention  and  use a  constant $\tau_0=1\times 10^{-13}$\,s  in the energy regime relevant here. 

\subsection{Quality of models}

Table~\ref{tab1} presents the average structural and energetic properties with the standard deviation computed over the 28 samples prepared following the procedure described in Sec.~\ref{model}. The quality of these models is established by comparing with similar \textit{a}-Si models generated following the same cooling schedule but using the mSW  potential~\cite{levesque_internal_2022}  as well as experimental results, when available. 

The final configurations for MTP-generated models show an energy per atom 0.14\,eV above that of the crystalline configuration at optimal density, as compared with the 0.135-0.205\,eV/atom from experiments~\cite{limbu_information-driven_2018} and 0.22\,eV/atom obtained with the mSW. At 2.23~g/cm$^3$, the density for MTP samples is slightly lower than the experimental value (2.28~g/cm$^3$), and similar to that produced with the mSW (2.2~g/cm$^3$). Regarding defects, the MTP samples contain $1.6\pm 0.6\%$ over-coordinated atoms (5-fold coordinated) and $0.6\pm 0.2\%$ under-coordinated defects (3-fold coordinated), a value similar to observed in   mSW samples, with $1.7\pm0.4\%$ over-coordinated and $0.7\pm 2\%$ undercoordinated atoms.

\begin{figure}
    \centering
    \includegraphics[width=1.0\linewidth]{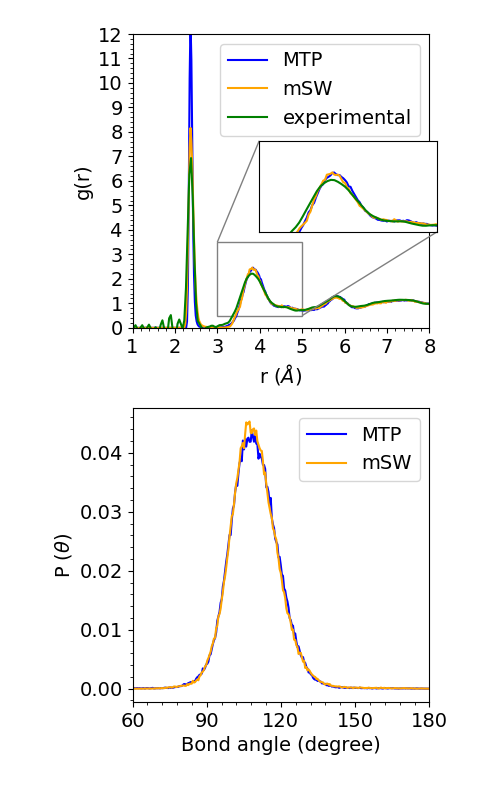}
    \caption{Top panel: The mean radial distribution function (RDF)  for the MTP ( total of 28 samples), the mSW (total of 200 samples) models and from diffraction experiments (EXP) \cite{laaziri_high_1999}. Bottom panel: The  mean angular distribution function  for the two sets of models.}
    \label{fig:rdf-adf}
\end{figure}

In Fig.~\ref{fig:rdf-adf} (top panel), we also compare the radial distribution for both MTP-  and mSW-generated samples with experimental data obtained using \textit{a}-Si membranes prepared by self-implantation~\cite{laaziri_high_1999}. We observe  good agreement between the various models and experiment, although, because models are evaluated at $T=0\,K$, the first peak, associated with the nearest neighbors, is narrower than for the experimental system, which shows a thermal broadening. While  experimental data are not available for the angular distribution, we see in the lower panel of Fig.~\ref{fig:rdf-adf} that both sets of models provide a similar distribution. 

Overall, the structural properties of the MTP-based models produced through a melt-and-quenched procedure are very similar to  available mSW-generated results and comparable to  experiments. Nevertheless, even considering the dispersion in the data, an argument can be made, taking into account all results presented in Table~\ref{tab1},  that MTP-based structures are less strained, show a similar number of defects, and are in overall slightly closer agreement with experiments than mSW-based models generated with the same simulation schedule using a potential specifically fitted to this phase of silicon. However, the difference for these structural properties is narrow and no differences in structural properties stand out between the two potentials. 

\begin{table*}
\caption{Properties of \textit{a}-Si prepared by melt and quench with  MTP (this article) and with mSW (\cite{levesque_internal_2022}). This includes the configurational energy, $E_c$, and the strain energy with respect to crystal ($ E_{\mathrm{strain}})$ per atom; the density $\rho$, the number of overcoordinated, fivefold, and undercoordinated, three-fold, atoms per 1000 atoms;  the number of TLSs in total and per 1000 atoms; and the number of independent configurations ($N_{\mathrm{samples}}$).
The properties are averaged for a sample size of 200 for mSW and 28 for MTP with uncertainty corresponding to the standard deviation. Experimental density is taken from Ref.~\cite{custer_density_1994} and experimental energy difference betwen crystalline and amorphous configuration from Ref.~\cite{limbu_information-driven_2018}.}

\centering
\begin{tabular}{c c c c}
 Sample & mSW & MTP & Experimental\\ [0.5ex] 
 \hline\hline
 $E_c$ (eV/at.)& $-3.078 \pm 0.004$  & $ -635.604 \pm 0.002$ & -\\ 
 \hline
 $ E_{\mathrm{strain}}$ (eV/at.) & $0.219 \pm 0.004$ & $0.140 \pm 0.002$ & 0.135 – 0.205\cite{limbu_information-driven_2018} \\
 \hline
$\rho$ (g/cm$^3$) & $2.200 \pm 0.006$ & $2.231\pm 0.001$ & 2.280\cite{custer_density_1994}\\
 \hline
 5-fold/1000 at. & $17 \pm 4$ & $ 16 \pm 6$ & -\\
 \hline
3-fold/1000 at. & $7 \pm 2$ & $ 6 \pm 2$ & -\\
 \hline
 TLSs (total)  & 390 & 117 & -\\ [1ex] 
 \hline
 TLSs/1000 at. & 1.95 & 4.18& -\\
 \hline
$N_{\mathrm{samples}}$ & 200 & 28 & - \\
 \label{tab1}
\end{tabular}
\end{table*}
 
\subsection{Two-level systems}

\begin{figure}[h]
    \centering
    \includegraphics[width=1.0\linewidth]{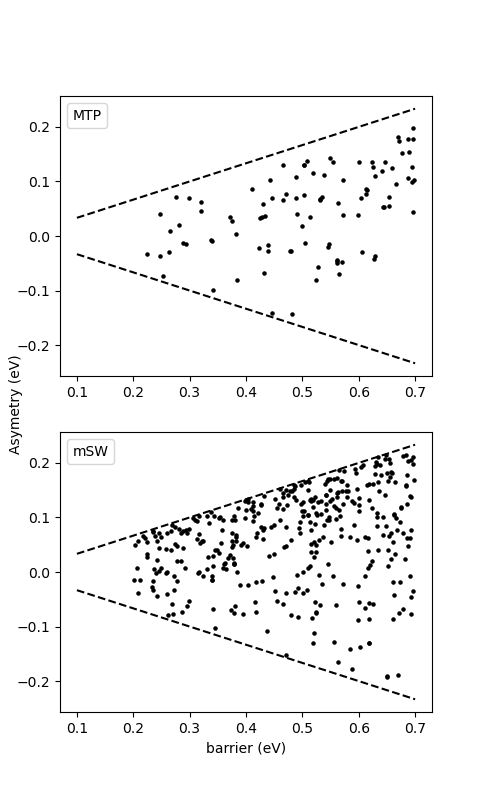}
    \caption{Energy asymmetry as a function of the forward energy barrier for events corresponding to the TLS definition criteria.  The doted lines delimit the cone of the acceptable asymmetry as a function of energy barrier. Top panel: results generated in this study; bottom panel, TLS generated using mSW and taken from Ref.~\cite{levesque_internal_2022}.}
    \label{fig:asym-barrier}
\end{figure}

We now turn to the characterization of the energy landscape, focusing  on the TLSs.  Out  of the 52\,000 events present in the catalog described in Sec. \ref{artn}, we identify 117 unique TLS ($4.18$ TLSs per 1000 atoms)  that respect the required conditions on energy and asymmetry: a barrier, $V$, between 0.2 and 0.7\,eV, which are at the origin of noise at a frequency around 50\,Hz for temperatures between 124\,K and 300\,K, the temperature at which the mirrors of GWD are kept. The energy asymmetry between the two connected minima, $\Theta$, such that $ \Theta \leq V/3$~\cite{damart_atomistic_2018}, is classified as a TLS as illustrated in Fig.~ \ref{fig:asym-barrier}. As indicated in Table~\ref{tab1}, the number of TLSs found here corresponds to a TLS density that is more than double that found with samples prepared using mSW (1.95 TLS/1000 atoms). (It is worth noting that despite a higher TLS density with the MTP, smaller TLSs statistics appear in the top panel of Fig.~ \ref{fig:asym-barrier} because  28 samples were studied, compared to 200 with the mSW potential.) 

Figure~\ref{fig:asym-barrier} (top panel) shows the energy asymmetry as a function of the energy barrier for the 117 TLSs measured from the initial state. Because each initial state is the end result of the extended quenching schedule presented above, we observe an asymmetry in the energy of the second states of TLSs, which are more often of higher energy, in agreement with previous observations~\cite{kallel2010}. This does not prevent, of course, some final states from having a lower energy than their initial counterpart, especially in the case of lower (less than 0.5\,eV) forward-energy barriers, \NM{as measured from the initial state of the simulation.}

The energy distribution of TLS is very similar to that obtained using mSW~\cite{levesque_internal_2022} (bottom panel). The mean barrier energy of the MTP samples is $0.51 \pm 0.13$\,eV, compared to $0.48 \pm 0.14$\,eV for the mSW samples.

\begin{figure}[h]
    \centering
    \includegraphics[width=1.0\linewidth]{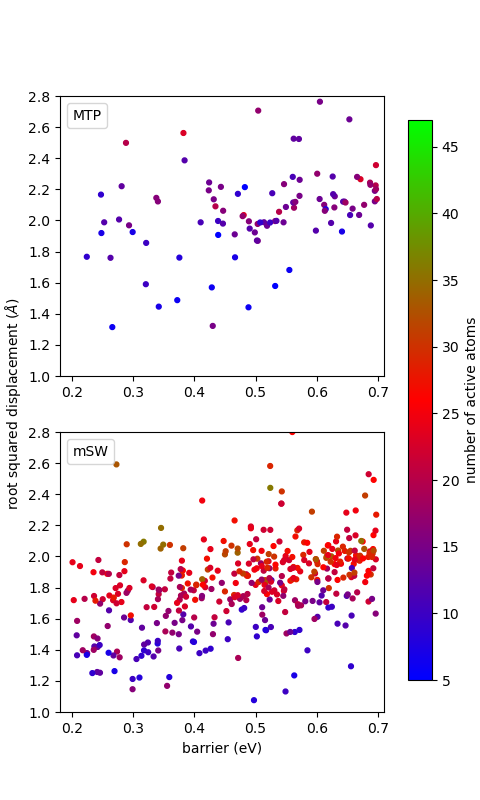}
    \caption{Mean square displacement between the initial and final states of TLSs as a function of the forward energy barrier. Symbols are colored according to the number of atoms moving by more than 0.1\,Å between the two states. Top panel: results generated in this study; bottom panel, TLS generated using mSW and taken from Ref.~\cite{levesque_internal_2022}.}
    \label{fig:dispersion}
\end{figure}

Figure~\ref{fig:dispersion} shows the root mean square displacement of the atoms as a function of the barrier energy. The number of atoms displaced by more than 0.1\,Å  between their initial and final states for each TLS is indicated by a color gradient. Again, the top panel shows the results for the samples obtained with the MTP while the bottom panel is for the TLSs previously found with the mSW potential. Although the average distance between minima is 2.05$\pm$0.25\,Å for TLSs generated in MTP samples, slightly longer than the  1.81$\pm$0.27\,Å for mSW, MTP TLSs tend to involve fewer atoms ($14\pm4$  vs. $22\pm7$ considering a range of one standard deviation). This suggests that the MTP energy landscape is somewhat smoother than mSW's: most atoms disturbed during activation therefore recover their original position, while those that move need often to make a bigger displacement to find a new stable position.

\begin{figure}[h]
    \centering
    \includegraphics[width=1.0\linewidth]{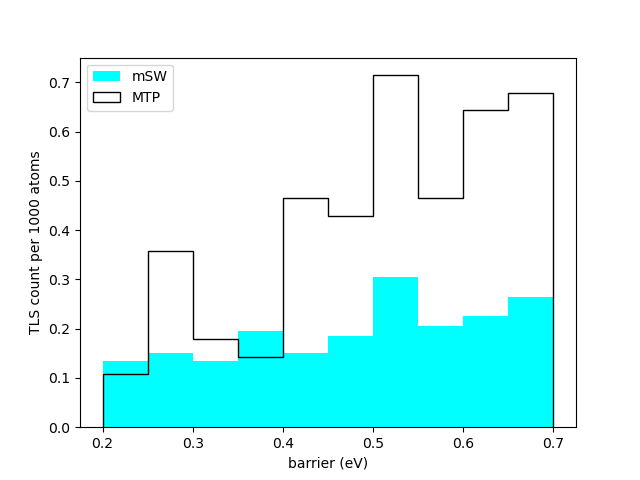}
    \caption{Histogram of the density of TLS events per 1000 atoms as a function of the forward energy barrier for MTP and mSW~\cite{levesque_internal_2022} results.}
    \label{hist-energ-tot}
\end{figure}

Finally, we present the histogram of the density of TLSs per 1000 atoms as a function of the forward energy barrier height, comparing MTP and mSW results (Fig.~\ref{hist-energ-tot}).  As indicated in the previous section, the overall density of MTP-based TLSs is higher than that for mSW models. The difference is not uniform, however: while the distribution is relatively flat for mSW samples, MTP-based TLSs of energy higher than 0.4\,eV are clearly more frequent. This bias toward higher forward energy barriers can be explained by the lower strain of the MTP models: with less energy stored in deformation, higher barriers need to be crossed to change state. Moreover, the more rigid network, observed when looking at the number of atoms involved in each event, might signify that local relaxation requires breaking bonds and that it cannot be accomplished simply with a slight diffusive-like shift of atoms.

\subsection{Categorization of the TLSs}

Following Ref.~\cite{levesque_internal_2022}, we categorize the TLSs according to the nature of the connectivity changes between the two linked minima: the first category of events represents bond-hop events, where one atom reduces its coordination to the benefit of a first-neighbor atom.
The second category defines events that involve a Wooten-Winer-Weaire (WWW) bond exchange mechanism~\cite{wooten_computer_1985}, a common occurrence in \textit{a}-Si~\cite{barkema_identication_1998,valiquette_energy}; this type of event involves two connected atoms exchanging a neighbor, while maintaining coordination for all atoms involved. We place all events that do not belong to either of these categories in a third class. To limit the effects of the cutoff, only atoms that move by at least 1\,\AA\, are considered for the definition of a bond hop or a  bond-exchange event.

\begin{figure}
    \centering
    \includegraphics[width=1.0\linewidth]{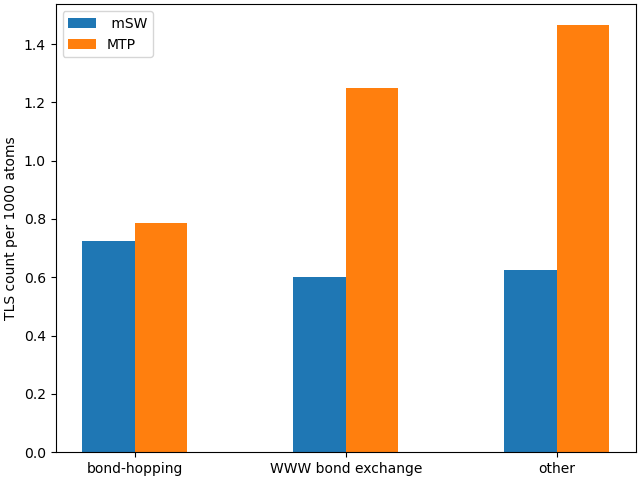}
    \caption{Average number of TLSs per 1000 atoms that belong to each of the three bond switching categories for the MTP-based (orange bars) and mSW-based models (blue bars).}
    \label{tls_dens}
\end{figure}

\begin{figure}[h!]
    \centering
    \includegraphics[width=1.0\linewidth]{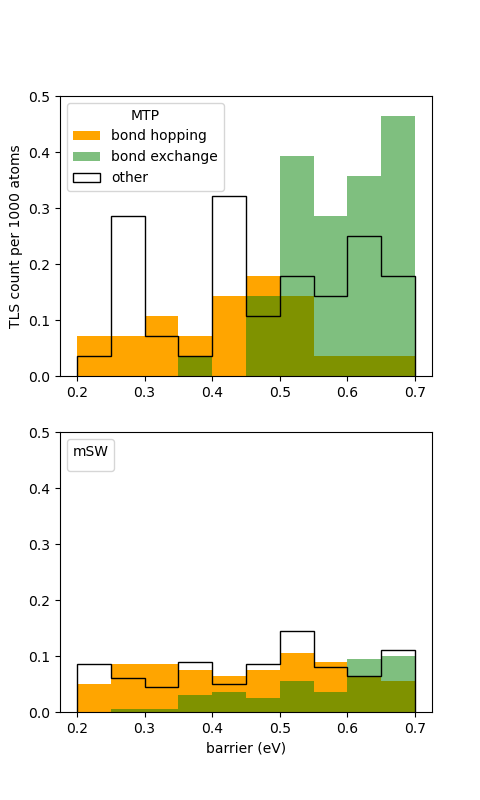}
    \caption{Density of TLS species per 1000 atoms in term of energy barrier for MTP (top) and mSW (bottom)\cite{levesque_internal_2022}.}
    \label{hist-energ-species}
\end{figure}

Among the 117 TLSs found, 19\% are bond-hopping events, 40\% are  bond-exchange and 41\% are in the other category. In Fig. \ref{tls_dens}, the density of TLSs per 1000 atoms for each type of TLSs for both mSW and MTP generated samples is illustrated. For bond-hopping events, a similar density of events is recorded, while there is a disparity between the models for both the bond-exchange events and the uncategorized events. As such, the difference in proportion of bond-hopping events compared to the other categories is only due to the higher number of  bond-exchange events and uncategorized events.  The contributions of the different types of TLSs to the density depending on barrier height are better illustrated in Fig. ~\ref{hist-energ-species}. In the top figure, the density of occurrences of each species is illustrated for the MTP generated samples. Lower barrier contributions to the density (below 0.5\,eV) are mainly from bond-hopping and uncategorized events, with contributions to higher energy barrier events originating mainly from  bond-exchange events. \NM{Overall, while the relatively small number of events leads to high fluctuations, bond-hopping events are concentrated in the low energy barrier fraction of TLs, bond-exchange events, in the high energy fraction, while the other category is distributed across the range.}
In the bottom figure, a similar histogram is illustrated, but only encompassing results from mSW generated samples. The difference in TLS density is smallest for the bond-hopping events,  as can be seen when comparing the data for the two types of samples at $0.038\pm 0.019$ with that of the "other" events at $0.1\pm 0.0$ and the  bond-exchange events at $0.1\pm 0.2$. At higher energy, the contribution of  bond-exchange events to the density is notably larger than what was estimated with the mSW potential, which is consistent with the previous observation that  bond-exchange events are much more common with MTP than with mSW. 

\subsection{Correlations between events}

Recent work have suggested that amorphous materials could display a much higher density of connected TLSs that would lead to the same experimental signature as rare isolated TLSs~\cite{blaber2025}. To assess whether this suggestion applies here, we examine whether the TLSs we generated are connected. 

While ARTn generated pure TLSs, composed of two minima directly connected by a common first-order saddle point, TLSs could be linked between themselves, forming a larger dissipative basin composed of multiple minima that would effectively decrease the number of independent TLSs~\cite{blaber2025}. To assess the proportion of these larger basins, we count the number of events with overlapping participating atoms, i.e. atoms in the same sample that move by at least 1\,Å. 

We find that most TLS recorded (99 events) in most samples (22 out of 28) are unique events, involving atoms that do not participate in other events.  Of the total 117 TLS found, 15\,\%, concentrated in 6 samples, are identified as involving active atoms recurring different events: five pairs of events are part of basins containing 2 events and 2 sets of four-overlapping-event basins; the 99 other events  are not connected to any other events and are isolated in their respective basin.

The weak spatial correlation between TLS is coherent with events which are local in nature, including single bond hop and  bond exchanges, and that can be analysed with the theory presented in Section~\ref{tls} and applied next.

\subsection{Loss angle calculation}

Using Eq.~\ref{loss_angle_eq}, we compute the loss angle $Q^{-1}$ as a function of temperature by summing the contribution of each TLS generated with the MTP. We  compare that sum  with measurements of the mechanical dissipation in \textit{a}-Si  from two samples of \textit{a}-Si grown by e-beam evaporation at 45°C and 200°C~\cite{liu_hydrogen-free_2014} as well as the sum obtained from the simulations with the mSW potential. The latter is  computed again for the current study as a factor 4 had been omitted in the calculation  of the elastic  modulus in  Ref.~\cite{levesque_internal_2022}; this leads to an overall downward shift of the curve with respect to  the one presented in that reference.

\begin{figure}[h!]
    \centering
    \includegraphics[width=1.0\linewidth]{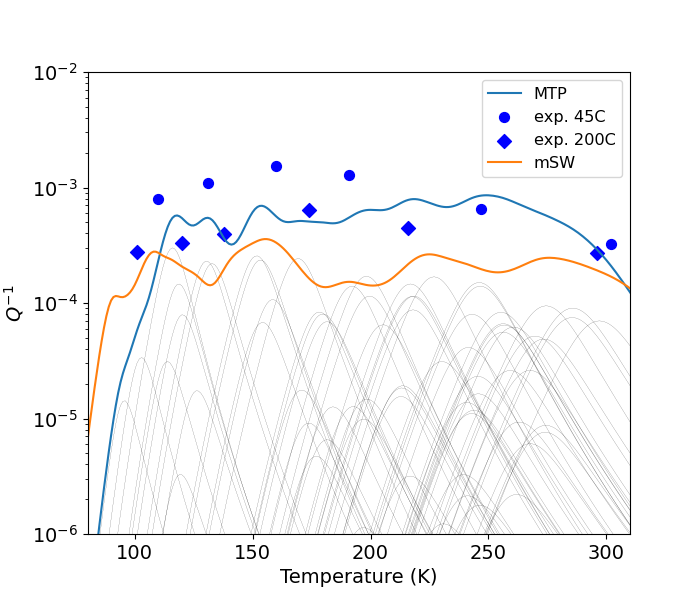}
    \caption{Loss angle contribution for MTP, mSW (recomputed with respect to Ref.~\cite{levesque_internal_2022}) and experimental results~\cite{liu_hydrogen-free_2014}. In grey, the contribution of individual TLSs of the MTP catalog to the mechanical noise.}
    \label{loss_figure}
\end{figure}

As illustrated in Fig.~\ref{loss_figure}, the noise computed from MTP data shows an excellent overlap with experimental results, suggesting that both the density  and the distribution  of  TLSs  computed with that potential are closer to what is observed experimentally than those computed using an mSW potential. 

\subsection{Validation of the MPT events}\label{mtp-results}

\NM{
To estimate the validity of the saddle points discussed above, TLS minima and saddle points obtained with MPT are reconverged using the mSW potential, as well as two MPT potentials with different levels of description fitted to the same Gaussian Approximation Potential (GAP), a DFT-based general-purpose ML potential for Si~\cite{Bartok2018}, by Morrow \textit{et al.}~\cite{Morrow2022}. More specifically, we consider here the indirect 16 (Ind16) and indirect 20 (Ind20) potentials of Ref.~\cite{Morrow2022}. This comparison with two potentials fitted to the same high-quality dataset allows for a better estimation of the intrinsic value of the results. }

\begin{figure}[tb]
    \centering
    \includegraphics[width=1.0\linewidth]{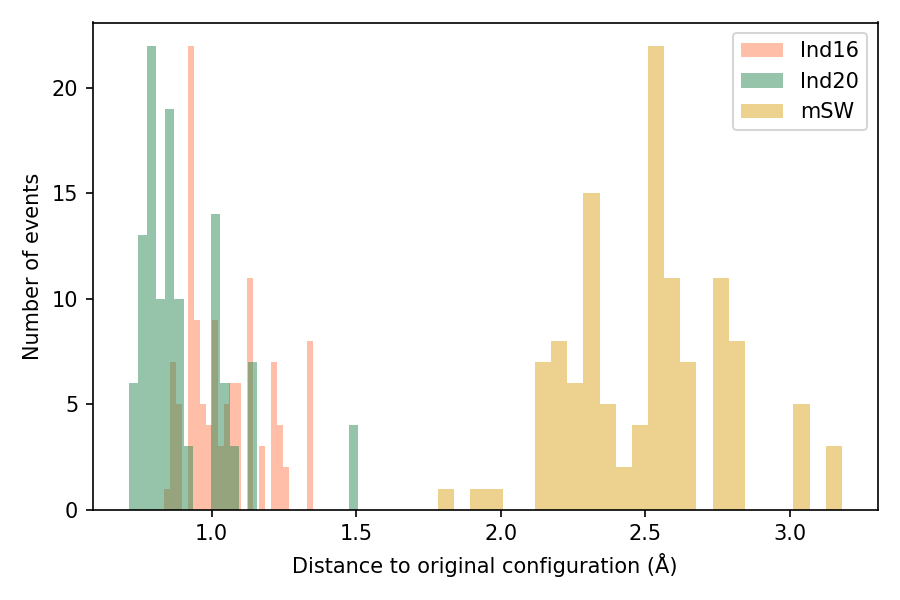}
    \caption{Displacement from minimum-energy configuration generated with MPT after relaxation with the empirical aSW as well as two ML potentials: Ind16 and Ind20.}
    \label{fig:rmin}
\end{figure}

\NM{Figure~\ref{fig:rmin} shows the displacement from the various minima obtained with MTP  when reminimized with the three potentials. While Ind16 and Ind20 show a displacement of around 1 Å away from the initial minima for the 1000-atom configurations, the mSW leads to significantly larger relaxations away from the MPT minima, suggesting that MPT minima are stable with mSW.} 

\begin{figure}[tb]
    \centering
    \includegraphics[width=1.0\linewidth]{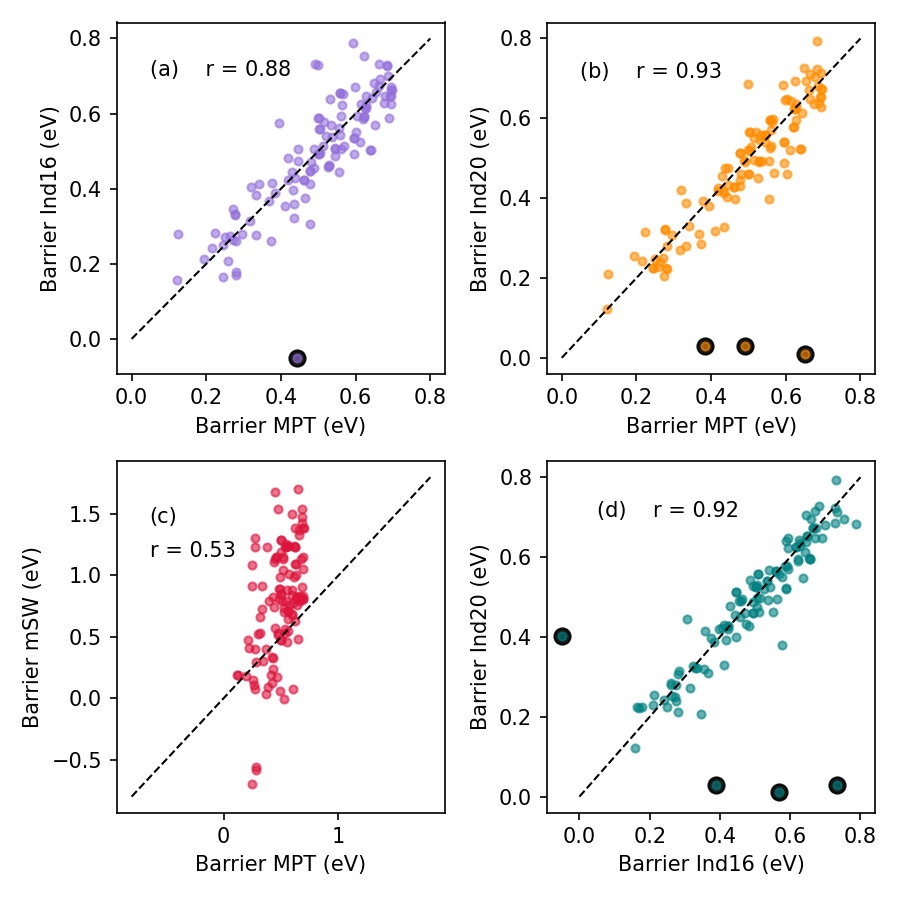}
    \caption{Correlations between the barrier height for MPT generated TSL saddle points and the saddle points found after a further convergence with (a) Ind16, (b) Ind20 and (c) aSW potentials. Panel (d) shows correlations between the barrier heights obtained with Ind16 and Ind20. R-values indicated in the panel indicated the regression coefficient.  Points surrounded by a black line are not included in the evaluation of the regression coefficient.}
    \label{fig:corsad}
\end{figure}

\begin{figure}[tb]
    \centering
    \includegraphics[width=1.0\linewidth]{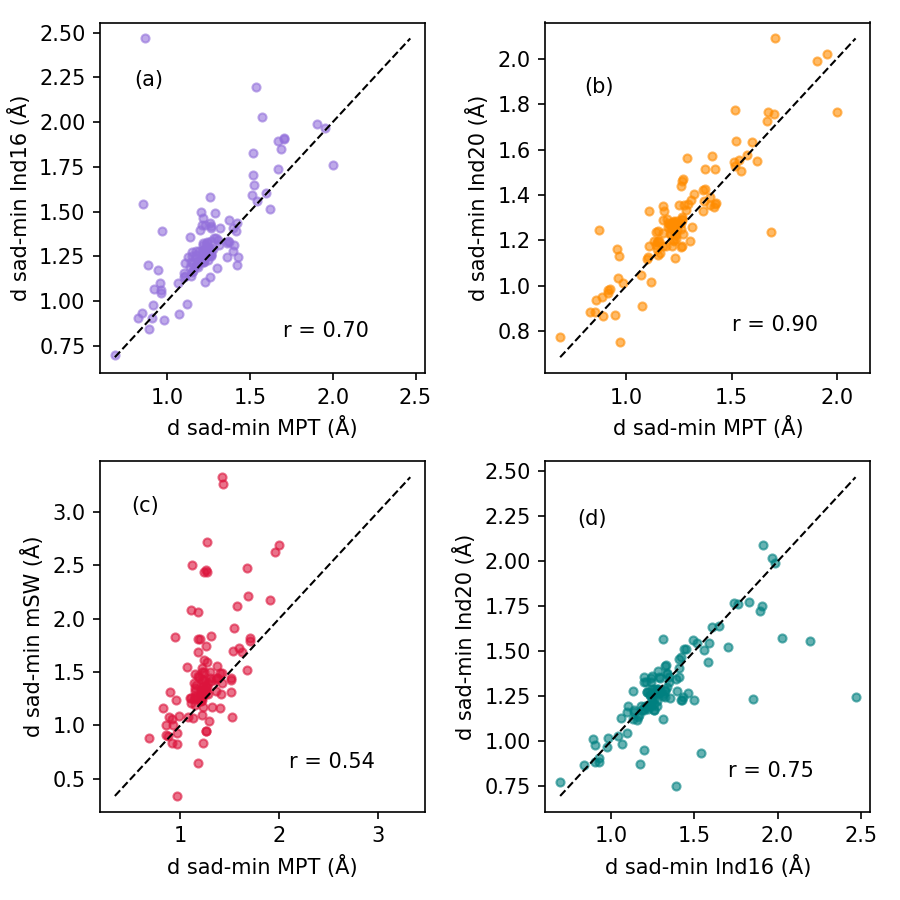}
    \caption{Correlations between the displacement between the saddle point and initial minimum for MPT generated TSL saddle points the displacement found after a further convergence with (a) Ind16, (b) Ind20 and (c) aSW potentials. Panel (d) correlations between the barrier heights obtained with Ind16 and Ind20. R-values indicated in the panel indicated the regression coefficient.  Points surrounded by a black line in Fig~\ref{fig:corsad} are not plotted nor included in the evaluation of the regression coefficient.}
    \label{fig:cordist}
\end{figure}

\NM{To assess the quality of saddle points, we plot the correlations for the saddle energy (Fig.~\ref{fig:corsad}) and the displacement from the initial minimum to the saddle point (Fig.~\ref{fig:cordist}) between the MPT structures discussed previously and those obtained after further convergence with Ind16, Ind20, and mSW. Looking at Fig.~\ref{fig:corsad}, we note that the majority of saddle points are reconverged to structures energetically close to the initial MTP structure, except for a handful of saddle points (one for Ind16 and a different set of three for Ind20) that relax into a minimum-energy state (barrier near zero eV). If we set aside these few points, we find a similar degree of correlation between the three ML potentials, with a slightly better agreement between Ind20 and MPT, even though Ind16 and Ind20 are fitted to the same data set. As expected from the comparisons made in the previous sections, the correlation is much lower when compared with mSW. Results are similar when we compare the displacement between the initial minimum and the saddle point, confirming the strong similarity in the structures converged by the three ML potentials and, in particular, between MPT and the richer Ind20, even though very few barriers are not reproduced by the three potentials.}

\section{Conclusions}

Twenty-eight samples of \textit{a}-Si were prepared by melt-and-quench using a recently developed DFT-based machine-learned moment tensor potential (MTP)~\cite{zongo_unified_2024}. The energy landscape of these samples was explored using ARTn to identify TLSs. Out of 52\,000 generated different events, 117 were found to correspond to TLSs that contribute to the mechanical dissipation mechanisms in the frequency and temperature range relevant to reproduce the noise observed in GWD.  

The properties of the samples and of the TLSs were then compared with samples generated in a similar fashion but using the empirical mSW potential ~\cite{levesque_internal_2022}, to measure and validate the importance of using DFT-based potentials in the study of activated mechanisms.  The use of MTP leads to amorphous models with slightly fewer defects than those generated with mSW, with a marginally better agreement on these structural properties compared to experiments. This is expected since mSW was  modified to address some of the shortcomings of SW potential in generating realistic models of \textit{a}-Si~\cite{vink_fitting_2001}. Differences are more important when looking at the energy landscape. In spite of similar defect concentration, MTP models show a TLS density twice that of mSW, with a dominance of compact complex events, while mSW favors simple bond-hopping mechanisms, with longer-range deformation. These changes are captured, at the macroscopic level, in the loss-angle contribution, with MTP presenting a better agreement with experiments than mSW.

While generally producing the right picture, these results underline the limits of empirical potentials for the specific description of atomistic mechanisms in disordered systems, even when they lead to reasonable average properties. \NM{Of course, there can still be variations between ML-fitted potentials; they can be significantly smaller than those between these and empirical potentials, when fitted to sufficiently large datasets, as shown here.}  

Getting the atomistic details right is essential to really understand phenomena such as dissipation caused by TLSs and develop efficient strategies to reduce it. \NM{With higher confidence in the modeling results regarding the structure and kinetics of amorphous silicon, future work will be able to provide more reliable answers to these specific questions.}

\section{Acknowledgments}

This research was supported in part by the NSERC Discovery program and  
generous computing time allocation  provided by Calcul Québec  and the Digital Research Alliance of Canada. The authors would like to thank Sjoerd Roorda and members of the LIGO Scientific Collaboration's Optics Working Group for useful discussions.

\bibliography{references_article.bib}
\bibliographystyle{ieeetr}
\end{document}